\documentclass[journal=cmatex,manuscript=article]{achemso}

\usepackage[version=3]{mhchem} 
\usepackage{siunitx}
\usepackage{hyperref}

\author{Jaime Dolado}
\affiliation{Departamento de F\'{\i}sica de Materiales, Universidad Complutense de Madrid, E-28040 Madrid, Spain}

\author{Ruth Mart\'{i}nez-Casado}
\affiliation{Departamento de F\'{\i}sica de Materiales, Universidad Complutense de Madrid, E-28040 Madrid, Spain}

\author{Pedro Hidalgo}
\affiliation{Departamento de F\'{\i}sica de Materiales, Universidad Complutense de Madrid, E-28040 Madrid, Spain}

\author{Rafael Gutierrez}
\affiliation{Institute for Materials Science, TU Dresden, 01062 Dresden, Germany}

\author{Arezoo Dianat}
\affiliation{Institute for Materials Science, TU Dresden, 01062 Dresden, Germany}

\author{Gianaurelio Cuniberti}
\affiliation{Institute for Materials Science, TU Dresden, 01062 Dresden, Germany}

\alsoaffiliation{Dresden Center for Computational Materials Science, TU Dresden, 01062 Dresden, Germany}

\alsoaffiliation{Center for Advancing Electronics Dresden, TU Dresden, 01062 Dresden, Germany}

\author{Francisco Dom\'{i}nguez-Adame}
\affiliation{Departamento de F\'{\i}sica de Materiales, Universidad Complutense de Madrid, E-28040 Madrid, Spain}

\author{Elena D\'{i}az}
\affiliation{Departamento de F\'{\i}sica de Materiales, Universidad Complutense de Madrid, E-28040 Madrid, Spain}

\author{Bianchi M\'{e}ndez}
\affiliation{Departamento de F\'{\i}sica de Materiales, Universidad Complutense de Madrid, E-28040 Madrid, Spain}
\email{bianchi@ucm.es}

\title[UV luminescence of germanates]%
  {Understanding the UV luminescence of zinc germanate: the role of native defects}

\abbreviations{PL,SEM}


\begin{document}

\begin{abstract}

Achieving efficient and stable ultraviolet emission is a challenging goal in optoelectronic devices. Herein, we investigate the UV luminescence of zinc germanate \ce{Zn2GeO4} microwires by means of photoluminescence measurements as a function of temperature and excitation conditions. The emitted UV light is composed of two bands (a broad one and a narrow one) associated with the native defects structure. In addition, with the aid of density functional theory (DFT) calculations, the energy positions of the electronic levels related to native defects in \ce{Zn2GeO4} have been calculated. In particular, our results support that zinc interstitials are the responsible for the narrow UV band, which is, in turn, split into two components with different temperature dependence behaviour. The origin of the two components is explained on the basis of the particular location of Zn$_\text{i}$ in the lattice and agrees with DFT calculations. Furthermore, a kinetic luminescence model is proposed to ascertain the temperature evolution of this UV emission. These results pave the way to exploit defect engineering in achieving functional optoelectronic devices to operate in the UV region. 

\vspace*{2cm}

\noindent Published in Acta Materialia. J.\ Dolado \emph{et al.} Acta Mater. \textbf{196}, 626--634 (2020).\\[3mm]
\noindent \url{https://doi.org/10.1016/j.actamat.2020.07.009}

\end{abstract}

\newpage

\section{Introduction}

Wide bandgap semiconductor oxides, due to their structural and electronic properties, have emerged as key materials for efficient ultraviolet (UV) absorption and/or emission while being transparent to the visible light. Light-matter interaction in the UV range is a challenge in a broad range of applications, such as solar-blind photodetectors \cite{Tien2018,Tak2019}, solid-state lighting \cite{He2015}, sensors for health or environmental monitoring \cite{Sang2013}, to name a few. For these purposes, \ce{ZnO} ($E_g=\SI{3.4}{\electronvolt}$), \ce{SnO2} ($E_g=\SI{3.1}{\electronvolt}$) and \ce{TiO2} ($E_g=\SI{3.5}{\electronvolt}$) have been intensively investigated in the last decades, but their bandgap energy $E_g$ limits somehow going deeper in the UV region. Novel ultra-wide bandgap materials, such as \ce{Ga2O3} \cite{Lopez2014,Zhao2015} or alternatively ternary oxides, can overcome to some extend this drawback, enabling even bandgap engineering with wider bandgaps \cite{Su2014}. In this sense, zinc germanate \ce{Zn2GeO4} with $E_g\sim\SI{4.5}{\electronvolt}$ at room temperature, has recently been envisaged as a promising semiconducting oxide due to its intrinsic physical properties in terms of good electronic conductivity, optical transparency or chemical stability \cite{Mizoguchi2011}. The interest in this ternary oxide has already fostered new ideas in the design of efficient phosphors \cite{Wang2013}, high-performance solar blind photodetectors \cite{Zhou2016} and batteries \cite{Yi2013} that incorporate \ce{Zn2GeO4} nanoparticles and/or nanowires as active elements. However, there is still a lack of understanding of the UV light absorption and emission processes in connection to their microstructure, including their native defects structure, which is crucial to fully exploit its potential in UV optoelectronic devices.

Luminescence from semiconducting oxides usually present broad visible emission bands often controlled by donor-acceptor pair (DAP) radiative recombinations, in which oxygen vacancies play the major role as donor levels while cation vacancies are responsible for the acceptor levels \cite{Maestre2004, Stichtenoth2008, Lopez2016}. Visible emission bands have also been reported in \ce{Zn2GeO4} and attributed to the above mentioned DAP recombination processes in a general way \cite{Liu2007,Pham2016}. Moreover, its luminescence has been found to be influenced by the synthesis route of the material, signaling that not only vacancies but also interstitial defects could play an active role in the optical response \cite{Tsai2013, Liu2007}. Also, the observation of both visible photoluminescence (PL) and UV cathodoluminescence (CL) bands from high quality microrods of \ce{Zn2GeO4} obtained by a thermal evaporation method \cite{Hidalgo2016} suggests that either UV or visible emission is feasible in \ce{Zn2GeO4} under certain excitation conditions, i.e. via electrons or photons. The above findings have not been yet found a satisfactory explanation, and therefore, an in-depth comprehension of the electronic states induced by specific native defects is required.
  
Indeed, one of the appealing features of wide bandgap semiconductor oxides is their potential to interact with UV light. However, UV emission is rather difficult to achieve in undoped oxides because native defects induce traps that quench near band edge transitions. One way of luminescence tailoring is by doping with optically active impurities, which provides emission at specific wavelengths depending on the impurity of choice. To this end, rare-earth ions are very suitable due to their well-defined intraionic levels transitions. However, doping with heavy ions is not always straightforward because of the low ions diffusivity in the host crystal, and ion implantation methods might be needed, as reported in \ce{Gd^3+} implanted \ce{Ga2O3} to achieve UV emission lines peaked at $\SI{313}{\nano\meter}$ \cite{Nogales2011}. Besides, other issue for effectively doping oxides is the location of the impurity in the crystalline lattice and its interaction with native point defects, which could alter the electronic states in the material. In complex oxides, such as ternary ones, multiple options of intrinsic point defects happen with the concomitant creation of electronic traps in the bandgap. Hence, they would provide a valuable way of tuning the luminescence emission with no need of importing foreign cations. In particular, the crystalline structure of \ce{Zn2GeO4}, built up by corner-shared \ce{ZnO4} and \ce{GeO4} tetrahedra arranged in rings of size enough to accommodate self-interstitial cations, would provide extra recombination pathways for excess carriers and be eventually responsible for the observed luminescence bands. To the best of our knowledge, a careful study of the PL emission as a function of photon energy excitation and the temperature evolution has not been fully addressed in \ce{Zn2GeO4}. Moreover, although some works have reported first-principle calculations of electronic states in \ce{Zn2GeO4} crystals \cite{Liu2013, Xie2015}, there is still a lack of a thorough description of the electronic levels related to native vacancies and Zn interstitials in this ternary oxide. Therefore, in order to get an accurate picture of the optical properties and to ascertain the origin of the UV emission bands of \ce{Zn2GeO4}, experimental and theoretical work is needed to correlate the nature of point defects with the observed luminescence properties. 

In this work, we provide a comprehensive study of the UV emission of \ce{Zn2GeO4} by considering experimental PL results combined with first-principle calculations and theoretical modeling of \ce{Zn2GeO4} micro- and nanowires. We have found that UV emission exhibits a composed nature, with several components that can be separately excited by selective excitation. The experimental results have been compared with density functional theory (DFT) calculations, which have led us to suggest a theoretical model based on rate equations for the origin of these emissions in connection to electronic levels related to vacancies V$_\text{O}$, V$_\text{Ge}$, V$_\text{Zn}$ and interstitial Zn$_\text{i}$ native defects. The results of this work would also be of major interest for applications since they pave the way to exploit defect engineering in achieving functional optoelectronic devices operating in the near and medium UV region. 
 
\section{Results and discussion}

\subsection{PL and PLE measurements}
\begin{figure}
    \centering
        \includegraphics[width=0.5
    \textwidth]{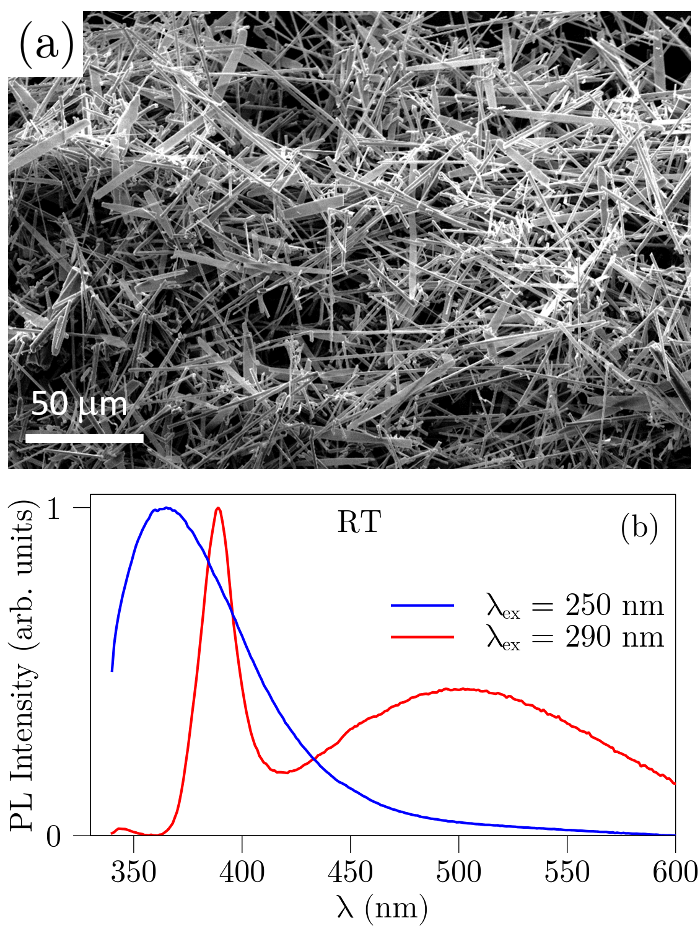}
        \caption{(a)~SEM image of \ce{Zn2GeO4} microstructures obtained by a thermal evaporation method. (b)~Room temperature PL spectra of \ce{Zn2GeO4} samples recorded at $\SI{250}{\nano\meter}$ ($\SI{4.96}{\electronvolt}$) and $\SI{290}{\nano\meter}$ ($\SI{4.27}{\electronvolt}$) excitation wavelengths.} 
        \label{fig1}
\end{figure}

Herein, \ce{Zn2GeO4} micro-wires have been synthesized by a catalyst-free thermal evaporation method as described in the methodology section. The microstructural analysis of the microwires has been assessed by Raman spectroscopy and X-ray diffraction measurements and it has been already reported \cite{Hidalgo2016, Dolado2020}. This procedure allows us to grow material of high crystalline quality. Figure~\ref{fig1}(a) shows a representative SEM image of the network of \ce{Zn2GeO4} structures under study. Figure \ref{fig1}(b) shows the room temperature (RT) PL spectra obtained under different excitation wavelengths, covering the range from $\SI{250}{\nano\meter}$ ($\SI{4.96}{\electronvolt}$), which is above the \ce{Zn2GeO4} bandgap ($E_g=\SI{4.5}{\electronvolt}$ at RT) up to $\SI{290}{\nano\meter}$ ($\SI{4.27}{\electronvolt}$), below the bandgap. These PL spectra reveal at least two UV emission bands centered roughly at $\SIrange[range-units = single,range-phrase=-]{340}{360}{\nano\meter}$ and $\SIrange[range-units = single,range-phrase=-]{370}{380}{\nano\meter}$, and another broad band peaked at $\SIrange[range-units = single,range-phrase=-]{500}{540}{\nano\meter}$. These bands are labelled as wide UV (W-UV), narrow UV (N-UV) and visible bands, respectively. For the sake of clarity, the PL spectra have been normalized to the N-UV emission band.

Most of the previous works on luminescence of \ce{Zn2GeO4} refer to room temperature experiments by exciting with energies under $E_g$, and only observe the visible band related to DAP transitions, while the detection of the UV emission has been scarcely reported \cite{Hidalgo2016}. Reported RT PL results also show that the maximum wavelength of the visible band varies from blue to green depending on the synthesis conditions, which is consistent with the interplay of several intrinsic defects in the luminescence mechanisms \cite{Tang2016, Liu2007}. Oxygen vacancies in semiconducting oxides have been claimed as the responsible for their rather good electronic conductivity, due to their behaviour as donors centres, in spite of their wide bandgap. In some oxides, as in \ce{Ga2O3}, some models proposed the formation of energy subbands within the bandgap providing delocalized electrons that contribute to the n-type electronic conductivity \cite{Binet1998}. The RT PL results of Figure~\ref{fig1} would support some partial conclusions about the light emission in \ce{Zn2GeO4}. Firstly, the broad shape of the W-UV and visible bands suggests that they are originated from recombinations between donors related to oxygen vacancies and either the valence band or acceptors levels, respectively. Secondly, Figure~\ref{fig1}(b) also shows that the emission bands strongly depend on the excitation wavelength. By exciting above the bandgap energy, only the W-UV band is produced while both the N-UV and the visible emission appear by exciting below the bandgap. This finding could explain the previous PL and CL results about getting UV emission under electron beam irradiation since CL provokes a higher excess carrier density that may recombine through all the available channels \cite{Hidalgo2016}. Lastly, the occurrence of the N-UV band suggests that another recombination donor center of a more localized nature than the oxygen vacancies is also playing a role. Liu \emph{et al.} have carried out electron paramagnetic resonance (EPR) measurements in \ce{Zn2GeO4} and proposed that, in addition to vacancies, Zn$_\text{i}$ interstitial defects could also be involved as donor centers in the DAP processes \cite{Liu2007}. 

\begin{figure}
    \centering
        \includegraphics[width=0.6\textwidth]{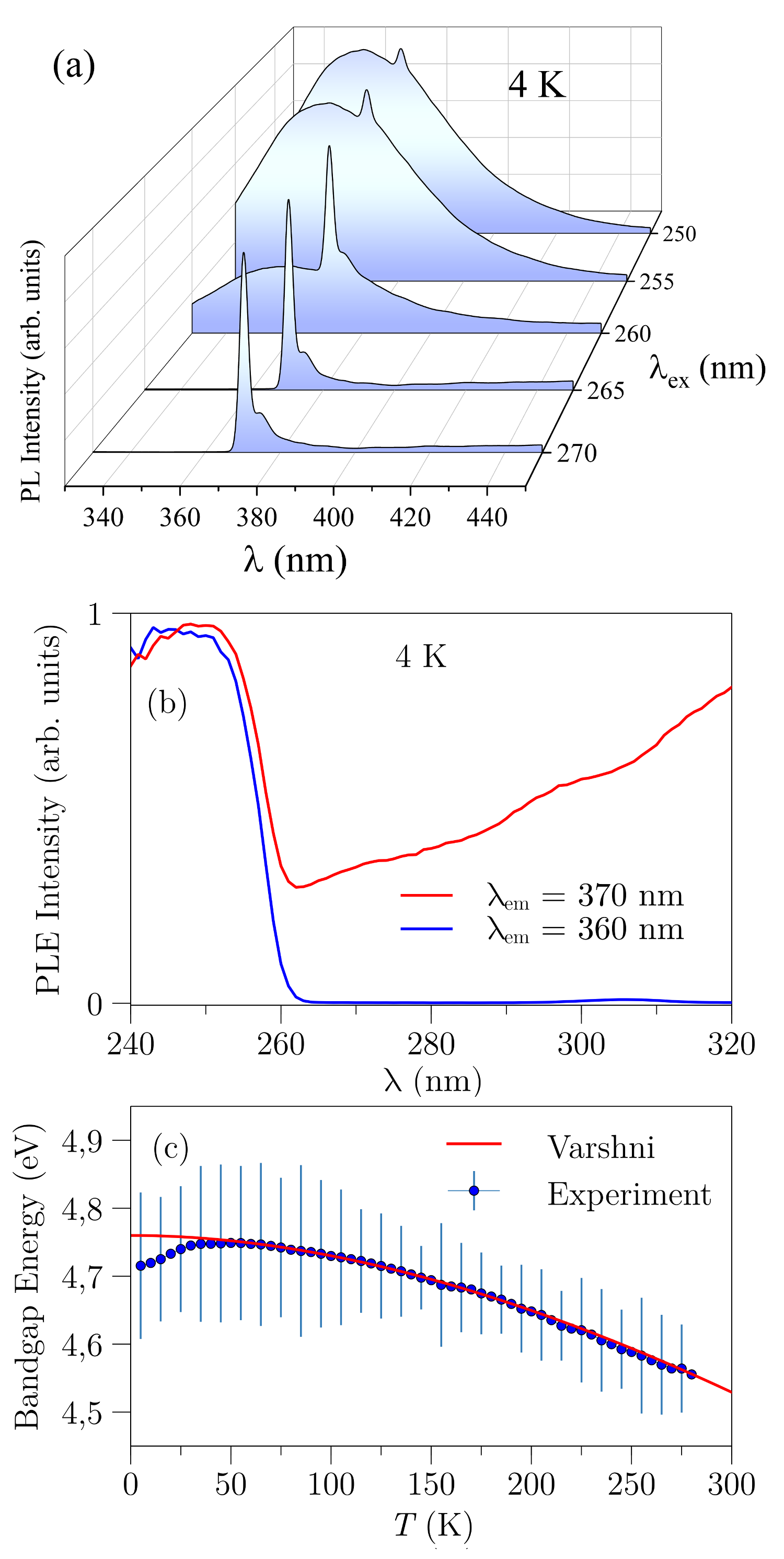}
        \caption{(a) PL spectra of \ce{Zn2GeO4} samples recorded at several excitation wavelengths $\lambda_\text{ex}$ recorded at $\SI{4}{\kelvin}$. (b) PL excitation spectra corresponding to the W-UV (blue line) and N-UV (red line) emissions. (c) Bandgap as a function of temperature. Blue circles correspond to experimental data obtained from PLE spectra from $\SIrange[range-phrase=\text{\ to\ }]{4}{300}{\kelvin}$. The red line shows the fitting to the Varshni equation~(\ref{eq1}). Experimental error bars are also displayed.} 
        \label{fig2}
\end{figure}

To further investigate the intriguing nature of the UV emission bands, low temperature PL and PL excitation (PLE) measurements have been performed. Figure \ref{fig2}(a) displays a series of PL spectra at $\SI{4}{\kelvin}$ recorded under different excitation wavelengths. The broad W-UV band ($\SI{360}{\nano\meter}$) is detected when exciting from $\SIrange[range-units = single,range-phrase=\text{\ to\ }]{250}{260}{\nano\meter}$, whereas the narrow N-UV band ($\SI{370}{\nano\meter}$) is observed for all the excitation wavelengths. No visible emission is detected at low temperature. In order to clarify the excitation conditions for the W-UV and N-UV bands, PL excitation (PLE) spectra have been recorded and shown in Figure \ref{fig2}(b). We can observe that the W-UV band is only excited by energies above the bandgap (blue line), while the N-UV can also be excited with lower photon energies (red line). In addition, the N-UV is actually split into two components ($\SI{370}{\nano\meter}$ and $\SI{375}{\nano\meter}$), clearly resolved when excitation is below the bandgap [see Figure \ref{fig2}(a)].

Therefore, from the above RT and $\SI{4}{\kelvin}$ PL results, we can conclude that different donor levels may be involved in the origin of the luminescence in \ce{Zn2GeO4}, offering alternative pathways for radiative transitions to either acceptor levels (leading to the visible emission) or to the valence band (UV emissions). Thus, besides the oxygen vacancies acting as donor centres, other native defects of composite nature would bring about the N-UV band.

PLE measurements indirectly provide an estimated value of the optical bandgap in a semiconductor by extrapolating the sharp decay at the shortest wavelengths to a linear function \cite{Masai2013}. We have recorded PLE spectra from $\SI{4}{\kelvin}$ to RT, from which experimental values of $E_g(T)$ are obtained and displayed in Figure~\ref{fig2}(c). Data were fitted by the semi-empirical relationship postulated by the Varshni equation\cite{Varshni67}
\begin{equation}
    E_g(T) = E_0 - \frac{\alpha T^2}{T + \beta} \ .
    \label{eq1}
\end{equation}
The temperature dependence of the bandgap is due to the effects of the thermal expansion of the lattice and the electron-phonon interaction. Usually, the balance between these effects produces a redshift of $E_g$ as temperature increases. The fitting of our data to Equation~\ref{eq1} yields $E_0=\SI{4.76}{\electronvolt}$, $\alpha=\SI{0.0036}{\electronvolt/\kelvin^2}$ and $\beta = \SI{1103.7}{\kelvin}$, although an anomalous behaviour is observed in the low temperature regime, from $\SIrange[range-phrase=\text{\ to\ }]{4}{40}{\kelvin}$. This anomaly has been reported in some semiconductors, such as \ce{CuInS2}, where it has been explained by the reduction of $d$-levels in the upper valence band due to thermal expansion and the competition with the electron-phonon interaction \cite{Hsu1988}.

\begin{figure}
    \centering
        \includegraphics[width=0.55\textwidth]{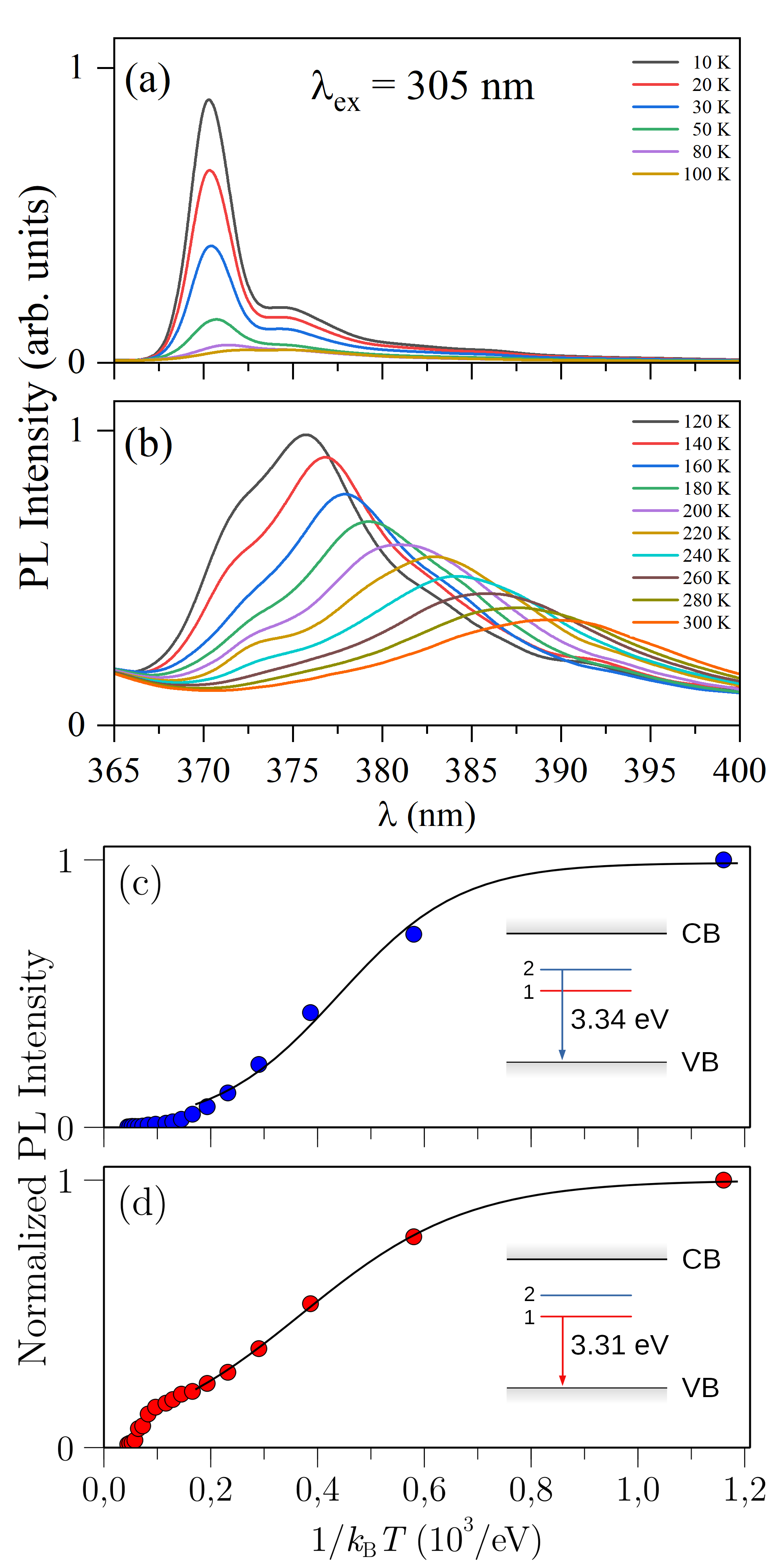}
    \caption{PL spectra as a function of the temperature obtained under $\SI{305}{\nano\meter}$ excitation wavelength, (a)~from $\SIrange[range-phrase=\text{\ to\ }]{10}{100}{\kelvin}$ and (b)~from $\SI{120}{\kelvin}$ to RT. PL intensity of the (c)~N-UV$_2$ and (d)~N-UV$_1$ components, respectively, as a function of $1/k_BT$. Experimental data (dots) are represented in comparison with a theoretical fitting (solid line) based on the energy band model explained below. Insets show the level diagrams illustrating the radiative transitions.} 
    \label{fig3}
\end{figure}

The monitoring of the two components of the N-UV band resolved in Figure~\ref{fig2}(a) as a function of the temperature has also been carried out. Figures~\ref{fig3}(a) and~\ref{fig3}(b) show a complete series of PL spectra under $\lambda_\text{exc} = \SI{305}{\nano\meter}$, from $\SI{10}{\kelvin}$ to RT. The selected $\lambda_\text{exc}$ allows us to study the nature and origin of the two components of the N-UV band as a function of temperature, without major contributions of the W-UV band. For the sake of clarity, PL curves have been separated into two plots, corresponding to low ($\SIrange[range-phrase=\text{\ to\ }]{10}{100}{\kelvin}$) and high ($\SIrange[range-phrase=\text{\ to\ }]{120}{300}{\kelvin}$) temperature range.
The analysis of the PL spectra as a function of the temperature gives rise to distinguish two temperature ranges. In the low temperature regime, from $\SIrange[range-phrase=\text{\ to\ }]{10}{100}{\kelvin}$ [see Figure~\ref{fig3}(a)], there is a marked drop of the intensity of the $\SI{370}{\nano\meter}$ band (N-UV$_2$) relative to $\SI{375}{\nano\meter}$ one (N-UV$_1$). On the other hand, in the $\SI{120}{\kelvin}$ to RT range [Figure~\ref{fig3}(b)], both components broaden in shape and experience a shift to lower energies. This different behaviour with temperature of the $\SI{370}{\nano\meter}$ and $\SI{375}{\nano\meter}$ bands leads us to consider that they could be originated from emission centers whose levels are close in energy. In some works, Zn$_\text{i}$ defects have been proposed to be involved in luminescence processes in \ce{ZnO} and \ce{Zn2GeO4} \cite{Wagner2013, Liu2007}. Our results are consistent with the hypothesis that the N-UV emission originates from Zn$_\text{i}$ related levels as well. The two components resolved in the N-UV band may be due to the two possible positions of the Zn$_\text{i}$ in the \ce{Zn2GeO4} crystalline lattice, which would modify the local environment of the defect. In particular, there are two types of six-member rings in the \ce{Zn2GeO4} lattice that could host Zn$_\text{i}$ at their centers. But in any case, both possible positions should be very similar since the energies of both centers are also similar, as we will show below from DFT calculations. 

Therefore, the temperature-dependent PL data ground the following luminescence mechanisms for the UV bands. Excitation above the bandgap would populate the V$_\mathrm{O}$ and Zn$_\text{i}$ related levels that will be further recombined with holes of the valence band leading to the UV luminescence, as it is observed in Figure~\ref{fig2}. In addition, the selective excitation at energies below the bandgap favors the N-UV emission at low temperature, suggesting a preferred recombination via Zn$_\text{i}$ related levels. The composed nature of the N-UV band would then be explained by the two possible local environment of Zn$_\text{i}$ defects. In the next subsections, we will show that these experimental findings are consistent with the first principle calculations results of defective \ce{Zn2GeO4}. Finally, a kinetic model for the luminescence process is proposed, providing further support to the relation of these native defects with the radiative recombination channels observed in PL spectra.

\begin{figure}
    \centering
        \includegraphics[width=0.9\textwidth]{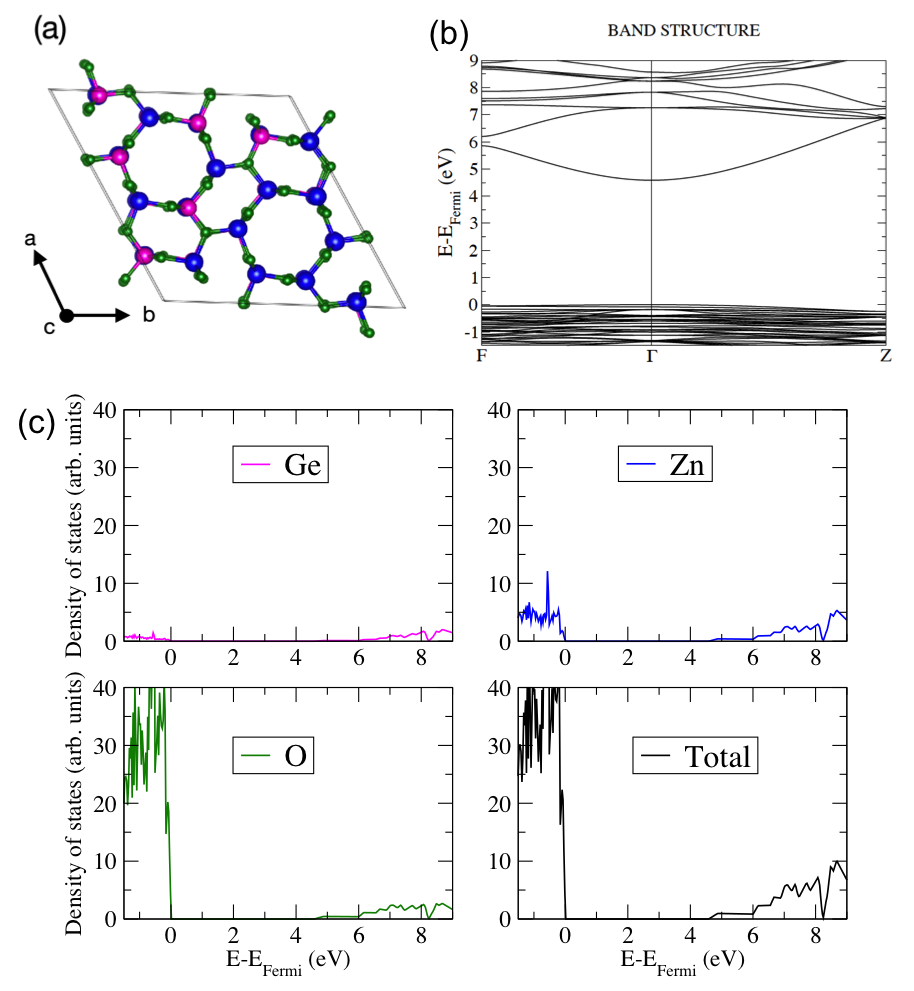}
        \caption{(a)~\ce{Zn2GeO4} unit cell view from \emph{c}-axis. Zn, Ge and O atoms are represented as blue, magenta and green circles, respectively. \textbf{a}, \textbf{b} and \textbf{c} correspond to the $x$, $y$ and $z$ axes, respectively. (b)~Band structure for the stoichiometric \ce{Zn2GeO4} and  (c)~DOS for the stoichiometric  \ce{Zn2GeO4}. Blue, magenta, and green lines are the projections on the Zn, Ge and O atoms, respectively. The black line shows the total DOS. } 
        \label{fig4}
\end{figure}

\subsection{First-principles calculations}

DFT calculations using the screened hybrid-exchange functional Heyd--Scuseria--Ernzerhof (HSE)~\cite{hse} have been performed in order to analyze the effect of the native defects in the electronic structure of \ce{Zn2GeO4}.  The optimized unit cell can be seen in Figure~\ref{fig4}(a), being the calculated lattice parameters ($a=b=\SI{14.27}{\angstrom}$ and $c=\SI{9.53}{\angstrom}$) compatible with the rhombohedral phase of \ce{Zn2GeO4} (JCPDS11–0687). The density of states (DOS) and band structure for the stoichiometric  \ce{Zn2GeO4} is shown in Figure~\ref{fig4}(b) and (c). The conduction band is mainly composed of Zn-$4s$, Ge-$4s$, and O-$2p$ orbitals, while the valence band is dominated by Zn-$3d$ and O-$2p$ orbitals. The calculated bandgap is $\SI{4.62}{\electronvolt}$, in excellent agreement with the experimental one at low temperature shown in Figure~\ref{fig1} ($\SI{4.76}{\electronvolt}$ at $T=\SI{4}{\kelvin}$). 
\begin{figure}
    \centering
        \includegraphics[width=1\textwidth]{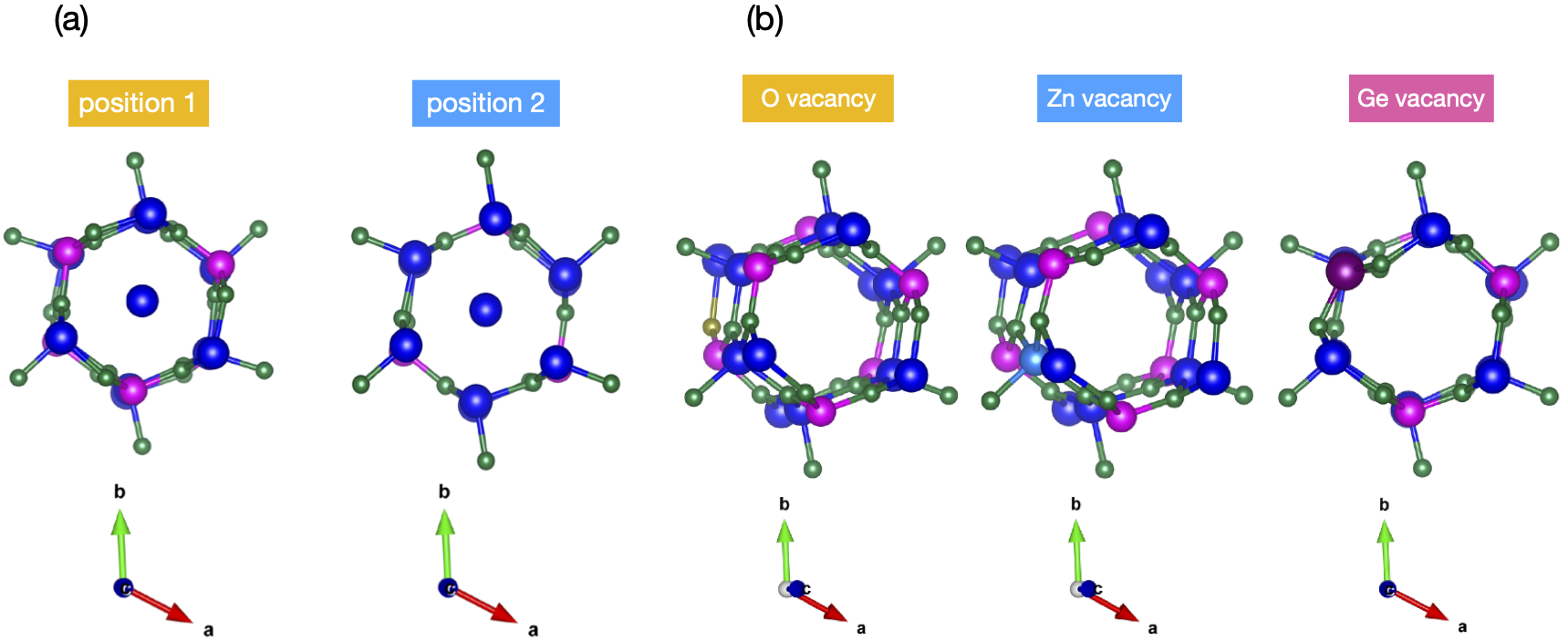}
        \caption{(a) Considered Zn$_{\text{i}}$ positions in \ce{Zn2GeO4}. Zn, Ge and O atoms are represented as blue, magenta and green circles, respectively. (b) Considered native defects in \ce{Zn2GeO4}. Zn, Ge, and O are represented as blue, magenta and green spheres. Orange, light blue and dark magenta represent the O, Zn, and Ge vacancies, respectively.  } 
        \label{fig5}
\end{figure}

\begin{figure}
    \centering
        \includegraphics[width=1
    \textwidth]{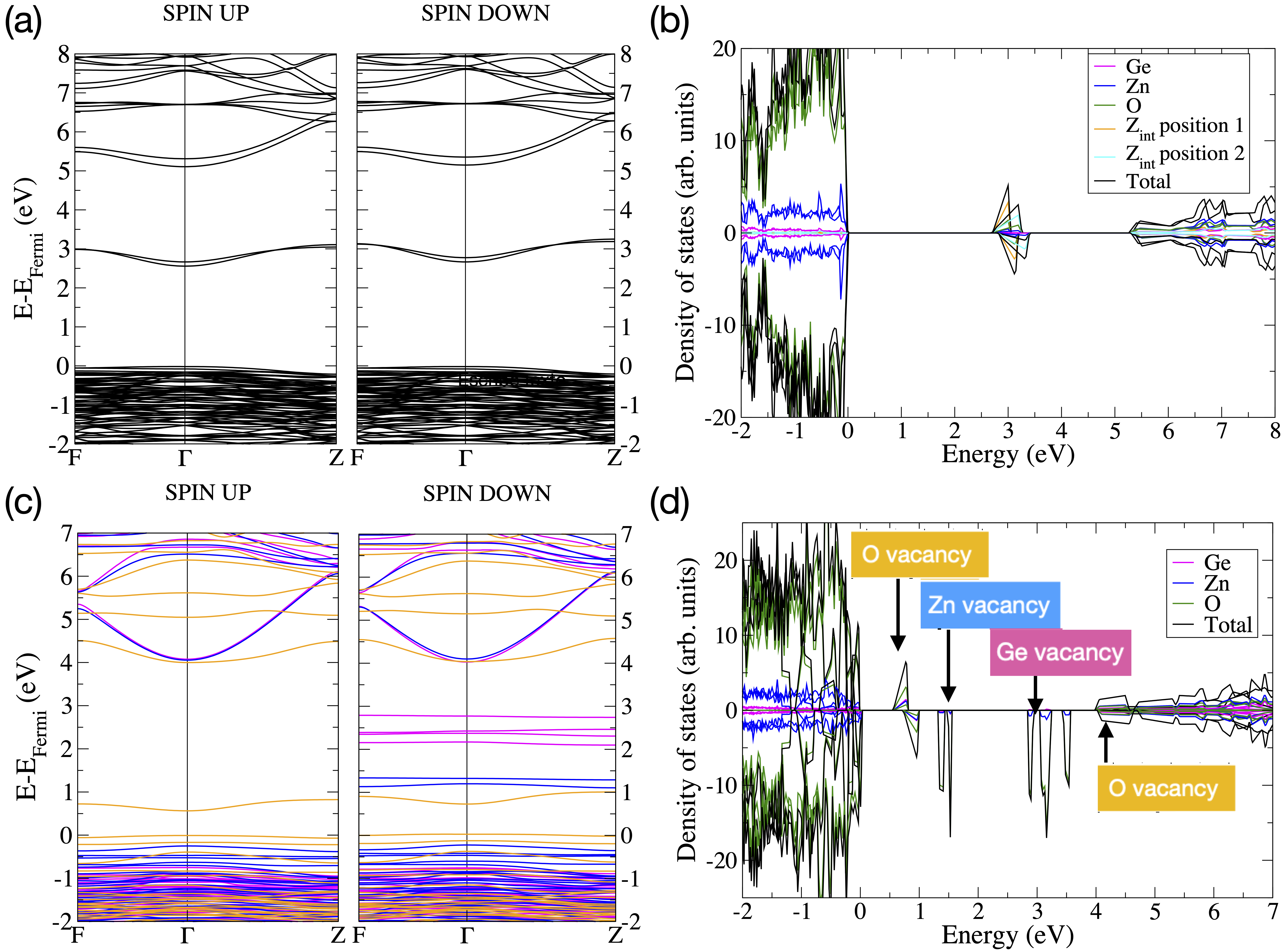}
        \caption{(a) Bands structure for Zn$_\text{i}$ in \ce{Zn2GeO4}. (b) DOS for Zn$_\text{i}$ in \ce{Zn2GeO4}. Blue, magenta and green lines are the projections on the Zn,Ge, O atoms, respectively. The black line shows the total DOS. Dopant-induced states are represented in orange (position 1) and cyan (position 2). (c) Band structure for O (orange lines), Zn (blue lines), and Ge (magenta lines) vacancies in \ce{Zn2GeO4}.  (d) DOS for O, Zn, and Ge vacancies in \ce{Zn2GeO4}. Blue, magenta, and green lines are the projections on the Zn, Ge, O atoms, respectively. The black line shows the total DOS. } 
        \label{fig6}
\end{figure}

Let us now move on to the results of defective \ce{Zn2GeO4}. DFT calculations of Zn$_{\text{i}}$ states are presented in Figures~\ref{fig5} and ~\ref{fig6}. The Zn$_\text{i}$ atom has been placed both at the center of the Zn-Ge ring (position 1) and Zn ring (position 2), as can be seen in Figure~\ref{fig5}(a).  The spin moment of the Zn$_{\text{i}}$ atom (0.1 $\mu_{B}$, $\mu_\text{B}$ is the Bohr magneton) leads to a DOS split into spin-up and spin-down states, and a dopant-induced state appears at the middle of the gap. The energies of the Zn$_{\text{i}}$ states, which are located at around $\SI{3}{\electronvolt}$ above from the valence band, are very similar in both structures, with an energy difference of around $\SI{0.2}{\electronvolt}$ between them [see Figure~\ref{fig6} (a) and (b)] . 

DFT results of O, Zn and Ge vacancy-related states are presented in Figures~\ref{fig5} and ~\ref{fig6}. The considered sites for the vacancy defects in \ce{Zn2GeO4} are shown in Figure~\ref{fig5}(b). The two unpaired electrons generated by the oxygen vacancy  lead to three different states in the gap. Two of them close to the valence band at $\SI{0.7}{\electronvolt}$ (spin up) and $\SI{0.9}{\electronvolt}$ (spin down), and another one very wide ($\SI{0.3}{\electronvolt}$ width)  and close to the conduction band at $\SI{4.2}{\electronvolt}$. According to the Mulliken population analysis, $0.3$ electrons go to the oxygen vacancy site, $0.7$ electrons are placed at the Ge contiguous to the vacancy and the rest are distributed over the remaining of oxygen atoms in the unit cell. In the case of the Zn vacancy, two spin down states appear at $\SI{1.4}{\electronvolt}$ and $\SI{1.6}{\electronvolt}$. These states are generated by the $0.92\,\mu_\text{B}$ spin moment of two oxygen atoms. Finally, the three unpaired electrons created by the Ge vacancy generate a spin moment of 0.98 $\mu_{B}$ in three oxygen atoms, which correspond to three gap states at $\SI{2.9}{\electronvolt}$, $\SI{3.2}{\electronvolt}$ and $\SI{3.6}{\electronvolt}$. The strong polarization that appears for both Ge and Zn vacancies have been previously reported in the literature~\cite{Xie2015}. The formation energy of the vacancies is calculated by using the expression
\begin{equation}
E_f = E_{\rm{def}} + \frac{1}{2} E_{\rm{X_{2}}} - E_{\rm{nodef}}\ ,
\label{eq_vac}
\end{equation}
where $E_{\rm{def}}$ and $E_{\rm{nodef}}$ are the total energies of the defective and non-defective structures, respectively. $E_{\rm{X_{2}}}$ represents the total energy of O$_2$ (X $=$ O), Zn$_2$ (X $=$ Zn) and \ce{Ge2} (X $=$ Ge). The obtained formation energies for the  O, Zn, and Ge vacancies are $\SI{6.4}{\electronvolt}$ , \SI{8.2}{\electronvolt}, and \SI{13.0}{\electronvolt}, respectively. For the Zn$_\text{i}$ defects, the formula used is
\begin{equation}
E_f =  E_{\rm{nodef}} + \,\frac{1}{2} E_{\rm{Zn_{2}}} - E_{\rm{def}}\ .
\label{eq_int}
\end{equation}
The resulting formation energies  for Zn$_\text{i}$ in position 1 and position 2 are \SI{2.2}{\electronvolt} and \SI{2.3}{\electronvolt}, respectively. The oxygen vacancy appears to be the most stable from the three considered ones, while the Zn$_\text{i}$ defects are more favourable. These calculations are in agreement with the luminescence measurements in \ce{Zn2GeO4} shown above and provide some hints about the electronic levels involved in the PL emission bands. The broad nature of the oxygen vacancies related levels is confirmed from DFT and support the presence of broad emission bands associated with oxygen vacancies in the UV or in the visible. Regarding the UV emissions, excitation with photons of energy above the bandgap populates the conduction band enough so that relaxation via transitions from V$_{\text{O}}$ and Zn$_{\text{i}}$ levels to the valence band lead to W-UV and N-UV luminescence bands, respectively. On the other hand, when exciting with energies below the bandgap, only the Zn$_{\text{i}}$-related emission is observed [see Figure~\ref{fig2}(b)] at low temperature. 

\subsection{Kinetics of the luminescent processes.}

\begin{figure}[htb]
\centerline{\includegraphics[width=0.4\columnwidth]{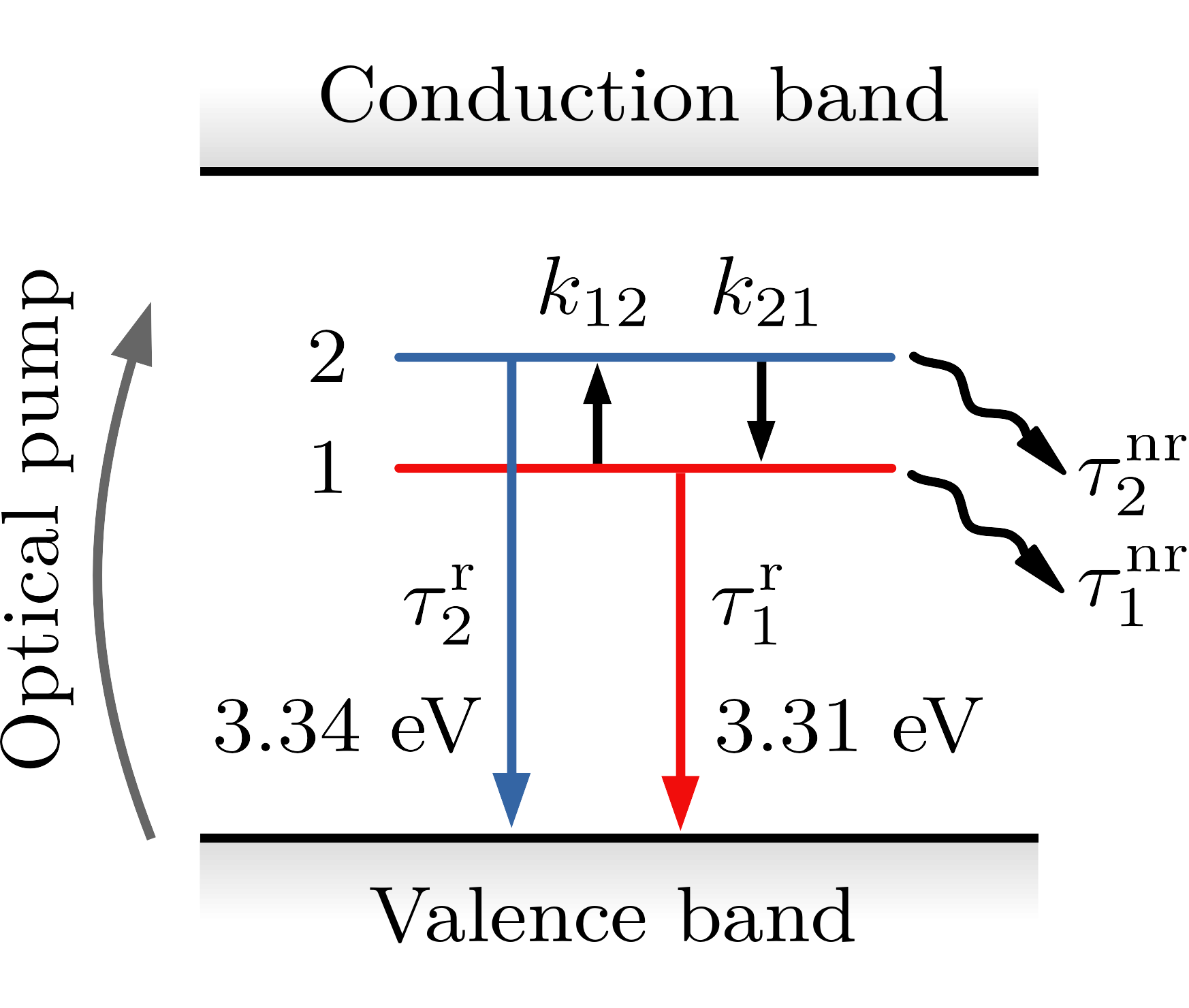}}
\caption{Energy band model and relevant transitions for interpreting the N-UV PL properties of native defects in \ce{Zn2GeO4}. $\tau_{i}^{\text{r}}$ and $\tau_{i}^{\text{nr}}$ with $i=1,2$ stand for the radiative and nonradiative decay time constants, respectively. We also sketch transitions between the two levels with nonradiative interlevel rates $k_{12}$ and $k_{21}$. At low temperature $k_{12}\simeq 0$.}
\label{fig_levels}
\end{figure}

The energy band model and the relevant transitions to explain the observed low temperature dependence of the N-UV PL intensity are shown in Figure~\ref{fig_levels}. After optical excitation with an energy below the bandgap, native defect levels become populated. Since the free-electron concentration $n$ in the conduction band is therefore negligible, the population of levels is only due to electrons generated by the optical pumping. Radiative and nonradiative channels deplete those levels. Let $E_i$, $\tau_{i}^{\text{r}}$ and $\tau_{i}^{\text{nr}}$ with $i=1,2$ be the energy, and the radiative and nonradiative decay time constants of such levels, respectively. To better understand the nature of the native defects, whether isolated or forming complexes, in this analysis we also consider transitions between the two levels $1$ and $2$ shown in Figure~\ref{fig_levels}, with nonradiative rates $k_{12}$ and $k_{21}$.

The concentration of occupied levels at a temperature $T$ and the total concentration of levels (occupied and empty)  will be denoted as $N_i(T)$ and $N_i^0$, respectively. According to the model, we can write down the following kinetic equations
\begin{equation}
\frac{dN_i}{dt}=-\frac{1}{\tau_i}\,N_{i}+g_i\left(N_i^0-N_i\right)+\sum_{j\neq i} 
\left(k_{ij}N_j-k_{ji}N_i\right)\ ,
\qquad
i,j=1,2\ ,
\label{eq_kinetics_01}
\end{equation}
where the dependence of $N_i$ on time and temperature is understood. The decay time constants are given as $1/\tau_i=1/\tau_{i}^{\text{r}}+1/\tau_{i}^{\text{nr}}$ and $g_i$ is the generation rate of level $i$ by the optical pump. Recently, Takahashi \emph{et al.} found that the selective introduction of interstitial Zn defects in Zn{$_2$}GeO{$_4$} yields a long-lasting PL emission~\cite{Takahashi10}. This finding agrees well with our observations and we thus take $\tau_{i}^{\text{r}}\gg \tau_{i}^{\text{nr}}$, namely $\tau_i \simeq \tau_{i}^{\text{nr}}$ in Equation~(\ref{eq_kinetics_01}).

Notice that in the absence of recombination mechanisms and interlevel transitions (i.e. $1/\tau_{i} \to 0$ and $k_{ij}\to 0$), solution to Equation~(\ref{eq_kinetics_01}) would be simply given as $N_i(t)=N_i^0\left[1-\exp(-g_it)\right]$, thus describing an exponential filling in time of the levels due to the optical pumping. It is most important to mention that we disregard electron capture on the empty levels in this model. The rate of electron capture is determined as $nC_i$, where $C_i\sim T^{3/2}$ \cite{Colbow68}. These processes would enter into the right-hand side of Equation~(\ref{eq_kinetics_01}) as a term of the form $nC_i\left(N_i^0-N_i\right)$. At low temperature and photon excitation energy below the bandgap, both $n$ and $C_i$ are small ($nC_i\ll g_i$) so we can neglect electron capture processes compared to the occupation of levels by optically generated electrons.

The interlevel rates $k_{12}$ and $k_{21}$ are not independent parameters since they must meet the principle of detailed balance
$k_{ij}/k_{ji}=\exp\left[\left(E_i-E_j\right)/k_\mathrm{B}T\right]$ ,
where $E_i$ is the energy of the electron level $i$ and $k_\mathrm{B}$ is the Boltzmann constant. Thus, in the absence of decay channels, the eventual electron distribution would account for the Boltzmann equilibrium distribution (nondegenerate electrons). According to DFT calculations and PL measurements, the energy difference of the most relevant levels in zinc germanate is $E_2-E_1\simeq\SI{30}{\milli\electronvolt}$. Therefore, at low temperature, namely $T\ll (E_2-E_1)/k_\text{B}\simeq \SI{360}{\kelvin}$, we can assume that $k_{12}\simeq 0$ and retain $k_{21}$ hereafter.

With these assumptions in mind, Equation~(\ref{eq_kinetics_01}) in steady state leads to
\begin{subequations}
\begin{align}
-\frac{1}{\tau_{1}^{\text{nr}}}\,N_{1}+g_1\left(N_1^0-N_1\right)+k_{21}N_2&=0\ ,
\label{eq_kinetics_03a}
\\
-\frac{1}{\tau_{2}^{\text{nr}}}\,N_{2}+g_2\left(N_2^0-N_2\right)-k_{21}N_2&=0\ .
\label{eq_kinetics_03b}
\end{align}
\label{eq_kinetics_03}%
\end{subequations}
Solving for $N_1$ and $N_2$, we can obtain the following temperature dependence of the intensity of both PL bands as $I_i=N_i/\tau_{i}^{\text{r}}$. Therefore, from Equations~(\ref{eq_kinetics_03}) we arrive at
\begin{subequations}
\begin{align}
I_1&=\frac{1}{1+1/(g_1\tau_{1}^{\text{nr}})}
\left(\frac{N_1^0}{\tau_{1}^{\text{r}}}
+\frac{\tau_{2}^{\text{r}}}{\tau_{1}^{\text{r}}}\,\frac{k_{21}}{g_1}\,I_2\right)\ ,
\label{eq_kinetics_04a}
\\
I_2&=\frac{N_2^0}{\tau_{2}^{\text{r}}}
\,\frac{1}{1+1/(g_2\tau_{2}^{\text{nr}})+k_{21}/g_2}\ .
\label{eq_kinetics_04b}
\end{align}
\label{eq_kinetics_04}%
\end{subequations}

According to Equation~(\ref{eq_kinetics_04b}), the intensity of the N-UV$_2$ band ($\SI{3.34}{\electronvolt}$) as a function of temperature can be cast in the form
\begin{equation}
I_2(T)=\frac{I_2^0}{1+A_2+B_2\exp(-E_{2a}/k_\mathrm{B}T)}\ ,
\label{I2-T}
\end{equation}
where $I_2^0$, $A_2$ and $B_2$ are temperature-independent parameters. Here $E_{2a}$ is an activation energy for nonradiative recombination of electrons in the upper level (level $2$ in Figure~\ref{fig_levels}). Figure~\ref{fig3}(c) shows the PL intensity of the N-UV$_2$ band ($\SI{3.34}{\electronvolt}$) as a function of temperature and the corresponding non-linear fit to Equation~(\ref{I2-T}) at low temperature ($T<\SI{80}{\kelvin}$). The fitting parameters are $A_2=0.0112$, $B_2=47.89$ and $E_{2a}=\SI{8.778}{\milli\electronvolt}$. According to the relation of these parameters with the microscopic parameters of the model appearing in Equation~(\ref{eq_kinetics_04b}), we conclude that $k_{21}$ is vanishingly small compared to the generation rate of the upper level by the optical pumping since $A_2\ll 1$. In other words, nonradiative transitions from the upper level to the lower level are irrelevant. This result gives further support to the claim that both deep levels are related to different and spatially separated point defects.

We now turn our attention to the PL intensity of the N-UV$_1$ band shown in Figure~\ref{fig3}(d). From Equation~(\ref{eq_kinetics_04a}) with $k_{21}=0$, the intensity of this band as a function of temperature is given as
\begin{equation}
I_1(T)=\frac{I_1^0}{1+B_1\exp(-E_{1a}/k_\mathrm{B}T)}\ ,
\label{I1-T}
\end{equation}
where $I_1^0$ and $B_1$ are temperature-independent parameters. Here $E_{1a}$ is an activation energy for nonradiative recombination of electrons in the lower level (level $1$ in Figure~\ref{fig_levels}). Recall that this expression only holds in the low temperature regime. The non-linear fitting of the experimental data at low temperature ($T<\SI{80}{\kelvin}$) yields $B_1=10.839$ and $E_{1a}=\SI{6.435}{\milli\electronvolt}$. The fact that $E_{1a}$ and $E_{2a}$ are not very different from each other is consistent with the similar surroundings of both native defects, as suggested from our DFT calculations.

\section{Conclusions}

The UV luminescence bands of \ce{Zn2GeO4} have been in-depth studied as a function of temperature and excitation energy conditions. The relationship between native defects and electronic states has been explored by means of DFT to ascertain the origin of the observed luminescence bands. A broad and a narrow UV emissions have been recorded and associated to oxygen vacancies-related levels and zinc interstitials-related levels, respectively. The particular features of the N-UV emission have motivated the proposal of a kinetic model to explain its temperature dependence. The results show that this emission is quite stable upon temperature changes and is the only dominant emission when exciting below the energy bandgap in the low temperature regime, up to $\SI{100}{\kelvin}$. The composite nature of this emission has been explained as originating from the two possible interstitial sites for Zn$_{\text{i}}$ within the Zn-Ge or Zn rings in the lattice structure. The energies of the related levels calculated by DFT nicely match the position of the PL maxima observed. Besides, the temperature dependence in the low temperature regime of both components has been theoretically modeled, yielding activation energies of about $\SIrange[range-units = single,range-phrase=-]{6}{8}{\milli\electronvolt}$.

\subsection{Methodology}
 
\emph{Experimental approach}. The studied samples have been obtained by a thermal evaporation method that uses a compacted pellet of a mixture of Zn, Ge and graphite as source material. The precursors used have been pure ZnO, Ge and carbon powders with a weight ratio 2:1:2 \cite{Hidalgo2016}. Then, a thermal annealing at $\SI{800}{\celsius}$ for 8 hours was conducted in an open tubular furnace not sealed under vacuum, leading to the growth of the \ce{Zn2GeO4} microstructures on the pellet. The microstructures were detached from the pellet and deposited on a silicon wafer for further characterization. The structural characterization has been carried out by grazing incidence X-ray diffraction (XRD) performed with a Philips X'Pert MRD Pro diffractometer. Raman measurements were conducted with the help of a Horiba Jobin Ybon LabRam Hr800 confocal microscope with excitation wavelength of $\SI{325}{\nano\meter}$. This optical microscope allows as well room temperature photoluminescence measurements. Finally, photoluminescence (PL) and PL excitation (PLE) spectra were acquired with an Edinburgh Instruments FLS1000 system at room temperature, exciting with a $\SI{450}{\watt}$ Xe lamp as excitation source.

\emph{Density Functional Theory calculations}. All DFT calculations have been performed using the CRYSTAL program~\cite{crystal}, in which the crystalline orbitals are expanded as a linear combination of atom-centered Gaussian orbitals, the basis set. The zinc, oxygen, and germanium ions are described using all-electron basis sets contracted as $s(8)\,p(64111)\,d(41)$, $s(8)\,p(621)\,d(1)$ and $s(9)\,p(76)\,d(511)\,p(31)$, respectively. 
Electronic exchange and correlation were approximated by using the Heyd–Scuseria–Ernzerhof (HSE) screened hybrid functional~\cite{hse}. {\bf In particular, the HSE06 functional with a mixing parameter of 0.25 has been used to describe the fundamental bandgap and electronic structure of ZnO, giving rise to underestimated bandgap values~\cite{hse-zno}. A mixing factor of 0.33 has also been tested in ZnO, yielding better results~\cite{clark2010}. In order to check the suitability of this mixing factor for \ce{Zn2GeO4}, we have performed a calculation using the HSE06 functional with a mixing factor of 0.33. The obtained bandgap with this mixing factor for stoichiometric \ce{Zn2GeO4} was of 5.2 eV, which is higher than our experimental value of 4.76 eV at 4K. However, the HSE06 with a mixing factor of 0.25 provides a bandgap of 4.62 eV, which is in excellent agreement with the experimental value and validates the considered model for \ce{Zn2GeO4}.} 
Integration over the reciprocal space was carried out using Monkhorst-Pack (MP) meshes of $ 6 \times 6 \times 6$. The self-consistent field (SCF) algorithm was set to converge at the point at which the change in energy was less than $10^{-7}$ Hartree per unit cell. The internal coordinates have been determined by minimization of the total energy within an iterative procedure based on the total energy gradient calculated with respect to the nuclear coordinates. Convergence was determined from the root-mean-square (rms) and the absolute value of the largest component of the forces. The thresholds for the maximum and the rms forces (the maximum and the rms atomic displacements) have been set to 0.00045 and 0.00030 (0.00180 and 0.0012) in atomic units. Geometry optimization was halted when all four conditions were satisﬁed simultaneously.

\begin{acknowledgement}

This work was supported by Ministerio de Ciencia, Innovaci\'{o}n y Universidades (Grants MAT2016-75955 and RTI2018-09195-B-I00) and Deutsche Forschungsgemeinschaft (Grant CU 44/47-1).

\end{acknowledgement}

\bibliography{bibliography}

\end{document}